\documentclass[12pt,draftclsnofoot, onecolumn]{IEEEtran}
\usepackage{graphicx,cite,epsfig,amssymb,amsmath}

\renewenvironment{cases}{\left\{\begin{array}[c]{ll}}{\end{array}\right.}

\begin{document}
\markboth{IEEE Trans. on WIRELESS COMMUNICATIONS, Vol. XX,
No. Y, Month 2010} {Ge etc.: Capacity of MIMO Cooperative Cellular Network \ldots}
\title{\mbox{}\vspace{0.40cm}\\
\textsc{Capacity Analysis of a Multi-Cell Multi-Antenna Cooperative Cellular Network with Co-Channel Interference}
\vspace{0.2cm}}
\author{\normalsize
Xiaohu Ge$^1$, \textit{Member, IEEE}, Kun Huang$^1$, Cheng-Xiang Wang$^2$, \textit{Senior Member, IEEE},
Xuemin Hong$^2$, \textit{Member, IEEE}\\
\vspace{0.70cm}
\small{
$^1$Department of Electronics and Information Engineering\\
Huazhong University of Science and Technology, Wuhan 430074, Hubei, P. R. China.\\
Email: xhge@mail.hust.edu.cn, kunhuang@smail.hust.edu.cn\\
\vspace{0.1cm}
$^2$Joint Research Institute for Signal and Image Processing, \\
School of Engineering \& Physical Sciences, \\
Heriot-Watt University, Edinburgh, EH14 4AS, UK.\\
Email: \{cheng-xiang.wang, x.hong\}@hw.ac.uk}\\
\thanks{\small{Submitted to IEEE Transactions on Wireless Communications.}}
\thanks{\small{The authors would like to acknowledge the support from the RCUK for the UK-China Science Bridges Project: R\&D on (B)4G Wireless Mobile Communications. X. Ge and K. Huang also acknowledge the support from the National Natural Science Foundation of China (NSFC) (Grant No.: 60872007), National 863 High Technology Program of China (Grant No.: 2009AA01Z239) and the Ministry of Science and Technology (MOST), China, International Science and Technology Collaboration Program (Grant No.: 0903). C.-X. Wang and X. Hong acknowledge the support from the Scottish Funding Council for the Joint Research Institute in Signal and Image Processing with the University of Edinburgh, which is a part of the Edinburgh Research Partnership in Engineering and Mathematics (ERPem).
}}}
\date{\today}
\renewcommand{\baselinestretch}{1.2}
\thispagestyle{empty}
\maketitle
\thispagestyle{empty}
\newpage
\setcounter{page}{1}\begin{abstract}

Characterization and modeling of co-channel interference is critical for the design and performance evaluation of realistic multi-cell cellular networks. In this paper, based on alpha stable processes, an analytical co-channel interference model is proposed for multi-cell multiple-input multi-output (MIMO) cellular networks. The impact of different channel parameters on the new interference model is analyzed numerically. Furthermore, the exact normalized downlink average capacity is derived for a multi-cell MIMO cellular network with co-channel interference. Moreover, the closed-form normalized downlink average capacity is derived for cell-edge users in the multi-cell multiple-input single-output (MISO) cooperative cellular network with co-channel interference. From the new co-channel interference model and capacity, the impact of cooperative antennas and base stations on cell-edge user performance in the multi-cell multi-antenna cellular network is investigated by numerical methods. Numerical results show that cooperative transmission can improve the capacity performance of multi-cell multi-antenna cooperative cellular networks, especially in a scenario with a high density of interfering base stations. The capacity performance gain is degraded with the increased number of cooperative antennas or base stations.

\end{abstract}
\newpage
\IEEEpeerreviewmaketitle
\vspace{-1cm}

\section{Introduction}

To achieve the high transmission data rate, multi-antenna technology has widely been adopted in the 3rd generation (3G) and 4th generation (4G) mobile communication systems [1]. The multi-antenna technology can improve the system capacity proportionally with the minimum number of antennas at the transmitter and receiver in a single cell communication system~[2]. However, the system capacity of multi-cell cellular networks is greatly degraded by the co-channel interference even with multiple antennas at the transmitters and receivers [3]--[7]. Therefore, a multi-cell multi-antenna cellular network is known as an interference-limited system [3], [4]. In this case, co-channel interference modeling and capacity analysis of multi-cell multi-antenna cellular networks are of great importance in the next generation mobile communication systems.

Numerous interference models have been proposed in the literature for wireless communication systems [8]--[18]. Kostic investigated the probability density function (PDF) of the interference caused by a single interfering transmitter over an interference channel with Nakagami-m fading and shadowing in wireless networks [8]. Liu and Haenggi explored the receiver interference PDF in  wireless sensor networks with some determinate interference node topologies, such as square, triangular, and hexagonal lattices interference node topologies in a two-dimensional plane [9]. References [10] and [11] studied the interference PDF performance of wireless networks with finite interfering transmitters based on wireless channels in [8]. However, the spatial distribution of interfering transmitters was not considered in [8]--[11]. Sousa proposed that the infinite aggregated interference from the homogeneous Poisson field of interferers can be modeled by an alpha stable distribution [12]. Furthermore, based on [12] an analytical expression for the instantaneous and second order distributions of the interference was presented in [13]. An interference model was provided for cognitive radio networks and the impact of channel fading and power control schemes on the interference model was analyzed in [14]. Salbaroli and Zanella investigated the interference characteristic function in a finite Poisson field of interferers in [15], [16]. The co-channel interference statistics in a Poisson field of interferers was derived from a unified framework in [17]. Gulati introduced a mathematical framework for the characterization of network interference in wireless networks in which interferers are scattered according to a spatial Poisson process and subject to path loss, shadowing, and \mbox{multi-path} fading [18]. Whereas, most models proposed in [8]--[18] were constrained to single antenna or single cell communication systems.

\hyphenation{whitening capacity random cooperative distributed transmitters communication interference methods}
For the capacity analysis of multi-antenna cellular networks, many studies have been carried out [19]--[31]. Foschini and Telatar carried out the initial research for the point-to-point multi-antenna communication system and indicated that the system capacity increases linearly with the minimum number of antennas in the transmitter and receiver over uncorrelated flat Rayleigh fading channels [2], [19]. Furthermore, in [20]--[22], the capacity was investigated for single-cell multi-user multiple-input multiple-output (MIMO) cellular networks over broadcast channels and some interference cancellation schemes, such as dirty paper coding (DPC), zero forcing (ZF), and block diagonalization (BD), were adopted to cancel the co-channel interference. Based on [22] and considering the finite equal noise interfering transmitters from adjacent cells, Shim improved the cellular network capacity by using a whitening filter for interference suppression at the receiver and a precoder using the interference-plus-noise covariance matrix for each transmitter [23]. Under a single cell communication system, Chiani developed an analytical framework to characterize the capacity of MIMO communication systems and derived the ergodic mutual information for MIMO systems in the presence of multiple MIMO co-channel interferers and noise [24]. In [25], [26], a centralized cooperative base station transmission scheme adopting joint transmission and cell selection scheme with co-channel interference was explored and the corresponding random capacity expressions were derived. By using the BD scheme to eliminate the inter-cell interference and dual decomposition scheme to optimize the power allocation, the random capacity of distributed cooperative BSs was investigated, and the simulation results showed that the distributed cooperative transmission scheme can achieve a better capacity performance than the centralized cooperative transmission scheme [27]. Without considering co-channel interference, the capacity of a distributed multi-cell zero-forcing beam-forming (ZFBF) in a multiple-input single-output (MISO) cellular network was analyzed in [28] and numerical results showed that the sum-rate per cell increases asymptotically with the number of users per cell. Furthermore, the new capacity expression in MIMO cellular networks extended from [28] was derived in [29]. For the cell-edge users, the capacity with beamforming cooperative transmission was investigated for the soft handover region in multi-cell MIMO cellular networks [30], [31].

However, in all the aforementioned capacity studies, only simple scenarios, such as a single cell with finite interfering transmitters, were considered and underlying channel models were limited to simple flat Rayleigh fading channels. Besides, the exact normalized average capacity of multi-cell MIMO cellular networks with co-channel interference has not been investigated. Moreover, more detailed investigation of the \mbox{analytical} co-channel interference model used for multi-cell multi-antenna cellular networks is surprisingly rare in the open literature. Because an analytical co-channel interference model for multi-cell multi-antenna cellular networks is not available, the co-channel interference was simply treated as noise in most previous work [26]--[28].

Motivated by the above gaps, in this paper we derive the exact downlink average capacity of multi-cell MIMO cellular network with co-channel interference. The contributions and novelties of this paper are summarized as follows.
\begin{enumerate}
\item We propose an analytical co-channel interference model for multi-cell MIMO cellular networks with the Poisson spatial distribution of interfering transmitters, taking into account fading and shadowing effects in wireless channels.
\item From the proposed co-channel interference model, we first derive the exact downlink average capacity of multi-cell MIMO cellular networks with co-channel interference.
\item The closed-form normalized downlink average capacity for cell-edge users in multi-cell MISO cooperative cellular networks with co-channel interference is derived for numerical analysis.
\item We study the normalized capacity of the multi-cell MISO cooperative cellular networks in great details and present some interesting observations.
\end{enumerate}

The remainder of this paper is outlined as follows. Section II proposes the new analytical co-channel interference model in multi-cell MIMO cellular networks.
In Section III, from the proposed co-channel interference model, the exact downlink average capacity of multi-cell MIMO cellular network is first derived.
Furthermore, a closed-form normalized downlink average capacity for cell-edge users in a multi-cell MISO cooperative cellular network with co-channel
interference is obtained. Numerical results and analysis are presented in Section IV. Finally, conclusions are drawn in Section V.


\section{Co-channel Interference Modeling And Performance Analysis}

\subsection{General Interference Model}
In this paper, interference analysis is focused on the downlink of cellular networks. To simplify the system model used for interference analysis, we only consider effective interferers in our general interference model. As shown in Fig. 1, there are only two types of nodes. One type of nodes is the signal receiver noted as user equipment (UE) with $N_{r}$ antennas and the other type of nodes is the interfering transmitter, i.e., base stations (BS) with $N_{t}$ antennas. Without loss of generality, the locations of all BSs are assumed to follow a Poisson spatial distribution in a two-dimensional infinite plane [32], [33]. Moreover, only one BS is assumed to exist in one cell. This general interference model can be used to describe the interference signals in multi-cell MIMO cellular networks.

\subsection{Interference Model of Multi-cell MIMO Cellular Networks}

In the aforementioned general interference model, as illustrated in Fig.1, every signal including the interference signal passes through an independent wireless channel, which means every signal is subject to independent path loss, Nakagami-m fading [34], and shadowing effect [35]. In general, the shadowing effect can be characterized by lognormal distributions. To simplify the calculation, the PDF of signal with shadowing effect can be approximated by Gamma distribution [33], [36]:
\[p(x) = \frac{1}{{\Gamma (\lambda )}}{(\frac{\lambda }{\Omega })^\lambda }{x^{\lambda  - 1}}{e^{ - \frac{\lambda }{\Omega }x}},x > 0   \tag{1a}
\]
with \\
\begin{equation}
\Gamma(\lambda)= \int_{0}^{\infty} t^{\lambda-1} e^{-t} \mathrm{d}t  \tag{1b}
\end{equation}
\begin{equation}
\Omega= P_r \sqrt{(\lambda+1)/\lambda}   \tag{1c}
\end{equation}
\begin{equation}
\lambda=1/(e^{(\sigma_{dB}/8.686)^{2}}-1)  \tag{1d}
\end{equation}
where $\sigma_{dB}$ is the shadowing spread parameter whose value usually ranges from 4 to 9 in practice and $P_r$ is the received average signal power at the receiver. Considering that the shadowing effect in the wireless channel is approximated by the Gamma distribution, the wireless channel with Gamma shadowing and Nakagami-m fading can be further approximated by the Generalized-K ($K_{G}$) distribution [33], [36].

Without loss of generality, we select one of the users $\mathrm{UE}_k$ $(k=1, \cdots, \infty)$ as the desired signal receiver in Fig. 1. This $\mathrm{UE}_k$ receives aggregated interference from the interfering transmitters, i.e., BSs, in the given region. The aggregated interference at $\mathrm{UE}_k$ can be expressed as follows
\[{P^{{R_X}}}
= \sum\limits_{b = 1}^\infty  {\frac{{{I_b}}}{{r_b^{{\sigma _r}}}}}
= \sum\limits_{b = 1}^\infty  {\frac{{\sum\limits_{i = 1}^{{N_r}} {\left( {{I_{b,i}}} \right)} }}{{r_b^{{\sigma _r}}}}}
= \sum\limits_{i = 1}^{{N_r}} {\left( {\sum\limits_{b = 1}^\infty  {\frac{{{I_{b,i}}}}{{r_b^{{\sigma _r}}}}} } \right)}
\tag{2}
\]
where $I_{b}$ is the interference signal power received by $\mathrm{UE}_k$ from the $\mathrm{BS}$ $b$ without the path loss effect, $I_{b,i}$ is the interference signal power received by the antenna $i$ of $\mathrm{UE}_k$ from the $\mathrm{BS}$ $b$ without the path loss effect, $r_{b}^{-\sigma_{r}}$ is the path loss variable with path loss coefficient $\sigma_{r}$ and path distance $r_{b}$ from the $\mathrm{BS}$ $b$ to the user. The spatial distribution of BSs is a Poisson distribution with a density parameter $\lambda_{BS}$. Thus, the distribution of the aggregated interference at $\mathrm{UE}_k$ is governed by an alpha stable distribution [17] and expressed by the characteristic function [16],[17].
\begin{equation}
\Phi_{P^{Rx}}(jw)=\mathrm{exp}\left(-|cw|^{\alpha} \left[1-j\mathrm{sign}(w)\mathrm{tan}\left(\frac{\pi\alpha}{2} \right) \right] \right)  \tag{3a}
\end{equation}
with \\
\begin{equation*}
\mathrm{sign}(w) =
\begin{cases}
  1,   &w > 0 \\
  0,   &w = 0 \\
  -1,  &w < 0
\end{cases}   \tag{3b}
\end{equation*}
\begin{equation}
\alpha=\frac{2}{\sigma_{r}}    \tag{3c}
\end{equation}
\begin{equation}
c=\sqrt [\alpha ] {\lambda_{BS}q\mathbb{E} \left({I_{b}^{\alpha} } \right)  }  \tag{3d}
\end{equation}
\begin{equation*}
q =
\begin{cases}
  \pi\Gamma \left(2-\alpha \right)\cos \left(\pi\alpha/2 \right)/ \left(1-\alpha\right),&\alpha \neq 1 \\
  \pi^{2}/2,   &\alpha = 1
\end{cases}   \tag{3e}
\end{equation*}
where $\mathbb{E}(\cdot)$ is the expectation operator, $\Gamma(\cdot)$ is the Gamma function which is defined by (1b), $\alpha$ is the characteristic exponent, and $c$ is the scale parameter.

In Fig. 1, an interfering transmitter, i.e., a BS, with $N_t$ antennas has $N_t$ interference sub-streams from the closed space source and every antenna of $\mathrm{UE}_k$ can receive $N_t$ interference sub-streams from a BS.
All of these interference signals coming from the same BS are assumed to
pass through the independent and identically distributed (i.i.d) Gamma shadowing processes and the
Nakagami-m fading processes. The transmission power of every antenna is assumed to
be equal and normalized to 1, i.e., $P_{ant}=1$. Therefore, in multi-cell MIMO cellular
networks, $I_{b}$ in (3d) is the sum of the interference signal power transmitted by
the $N_t$ interference sub-streams from the same $\mathrm{BS}$ $b$ without the path loss effect. Furthermore, $I_{b}$
is expressed by
\[{I_b} = {P_{ant}}{w_b}\left( {\sum\limits_{i = 1}^{{N_r}} {\sum\limits_{j = 1}^{{N_t}} {{{\left| {{z_{b,i,j}}} \right|}^2}} } } \right) = {w_b}\left( {\sum\limits_{i = 1}^{{N_r}} {\sum\limits_{j = 1}^{{N_t}} {{{\left| {{z_{b,i,j}}} \right|}^2}} } } \right)  \tag{4}
\]
where $w_b$ is a random variable of Gamma shadowing process,
which corresponds to the signal passing through the Gamma shadowing channel
from the $\mathrm{BS}$ $b$, and $z_{b,i,j}$ is the random variable of Nakagami-m fading process,
 which corresponds to the signal passing through the Nakagami-m fading channel from
 the transmitting antenna $j$ of $\mathrm{BS}$ $b$ to the receiving antenna $i$ of $\mathrm{UE}_k$.

Because the wireless channel with Gamma shadowing and Nakagami-m fading can be approximated by the $K_G$ distribution, the PDF of $I_b$ can be further derived as follows based on the Generalized-K random process theory [11], [36]:
\[{f_I}(y) = \frac{{2{{\left( {\frac{{m\lambda }}{\Omega }} \right)}^{\frac{{{N_t}{N_r}m + \lambda }}{2}}}}}{{\Gamma ({N_t}{N_r}m)\Gamma (\lambda )}}{y^{\frac{{{N_t}{N_r}m + \lambda  - 2}}{2}}}{K_{\lambda  - {N_t}{N_r}m}}\left( {2\sqrt {\frac{{m\lambda y}}{\Omega }} } \right)
\tag{5}
\]
where  $m$ is the Nakagami shaping factor, $K_v(\cdot)$ is the modified Bessel function of the second kind with order $v$, $\Omega$ and $\lambda$ are defined in (1c) and (1d), respectively.

To obtain the closed-form PDF of
(3a) for further performance analysis, a closed-form of (3d) should be first derived. Therefore, from (5) the
following transform parameter $\gamma$ is derived accounting for the transformation in [37]
\[\gamma=c^{\alpha} = \lambda_{BS}q\left( \frac{m \lambda}{\Omega} \right)^{-\alpha}
\frac{\Gamma(\lambda+\alpha)\Gamma(N_t N_r m + \alpha)}{\Gamma(N_t N_r m)
\Gamma(\lambda)}.    \tag{6}
\]

Furthermore, we substitute (6) into (3d) and perform the inverse Fourier
transform on the characteristic function of (3a). Ultimately a new PDF expression of the aggregated interference $P^{{R_X}}$
at the user in the multi-cell MIMO cellular networks is derived as
\begin{equation}
f_{P^{Rx}}(y)=\frac{1}{2\pi}\int_{-\infty}^{+\infty}
\Phi_{P^{Rx}}(j w) \mathrm{exp} (-2 \pi j w y)\mathrm{d}w    \tag{7a}
\end{equation}
where $\Phi_{P^{Rx}}(jw)$ is given by (3a) with \\
\begin{equation}
c = \left(\lambda_{BS}q\left( \frac{m \lambda}{\Omega} \right)^{-\alpha}
\frac{\Gamma(\lambda+\alpha)\Gamma(N_t N_r m + \alpha)}{\Gamma(N_t N_r m)
\Gamma(\lambda)}\right)^{\frac{1}{\alpha}}.  \tag{7b}
\end{equation}

When the path loss coefficient is
configured as $\sigma_r=4$ which corresponds to the urban macro-cell with rich scattering
environment [35], an analytical interference model is given by
\begin{equation}
f_{P^{Rx}}(y) = \sqrt{\gamma^{2}/2\pi}\frac{e^{-\gamma^{2}/2y}}{y^{3/2}}, y>0   \tag{8a}
\end{equation}
with \\
\[\gamma  = {\lambda _{BS}}\sqrt 2 \pi \Gamma (\frac{3}{2}){\left( {\frac{{m\lambda }}{\Omega }} \right)^{ - \alpha }}\frac{{\Gamma (\lambda  + \frac{1}{2})\Gamma ({N_t}{N_r}m + \frac{1}{2})}}{{\Gamma ({N_t}{N_r}m)\Gamma (\lambda )}}
\tag{8b}
\]
where $\Gamma(\cdot)$, $\Omega$ and $\lambda$ are denoted by (1b), (1c) and (1d), respectively.

\subsection{Performance Analysis}
Based on the proposed new interference model, some performance evaluations can be
numerically analyzed in detail. In the following analysis, some parameters of the
interference model in Fig. 1 are configured as default values: $\sigma_{dB}=6$,
$m=1$, $\sigma_{r}=4$, $N_t=4$, $N_{r}=2$, $\lambda_{BS}=1/(\pi \times500^{2})$, and $P_{r}=1$. In the previous work [38], Gaussian interference models were used. Based on the default
parameters, we compare the PDF of the proposed interference model with the PDF of
the conventional Gaussian interference model in Fig. 2. It is shown that
the PDF of the proposed interference model has obvious heavy tail characteristic
compared with the PDF of Gaussian interference model. The heavy tail characteristic
indicates that the small probability event, i.e., the rare event, has non-ignorable impact on
the alpha stable distribution. Therefore, the aggregate interference in multi-cell MIMO cellular networks can be easily dominated by individual high-power interfering signals.


Furthermore, we analyze the impact of some parameters on the proposed interference model in Figs. 3--8.
Fig. 3 shows that the probability of instantaneous interference power falling into the range from 0 to $3\times10^{-10}$ watt increases with the increase of the shadowing spread parameter $\sigma_{dB}$. But in the tail part, i.e., when the instantaneous interference power exceeds $3\times10^{-10}$ watt, the probability of the instantaneous interference power decreases with the increase of $\sigma_{dB}$. In Fig. 4, the probability of instantaneous interference power within the range from 0 to $2.5\times10^{-11}$ watt decreases with the increase of the BS density parameter $\lambda_{BS}$. When the instantaneous interference power exceeds $2.5\times10^{-10}$ watt, the probability of instantaneous interference power increases with the increase of $\lambda_{BS}$. Fig. 5 illustrates when the instantaneous interference power is less than $0.5\times10^{-9}$ watt, the PDF increases with the increase of the path loss coefficient parameter $\sigma_{r}$. The probability of instantaneous interference power gradually decreases with the increase of $\sigma_{r}$ when the instantaneous interference power is larger than $0.5\times10^{-9}$ watt. Fig. 6 demonstrates that the probability of instantaneous interference power decreases with the increase of the number of transmission antennas $N_t$ per interfering transmitter when the instantaneous interference power is less than $0.8\times10^{-10}$ watt. If the instantaneous interference power exceeds $0.8\times10^{-10}$ watt, the trend is reversed. In Fig. 7, the probability of instantaneous interference power decreases with the increase of the number of receiving antennas $N_r$ per user when the instantaneous interference power is smaller than $1.8\times10^{-10}$ watt. After this turning point, the probability of instantaneous interference power gradually increases with the increase of $N_r$. Based on Figs. 3--7 and the alpha stable process theory [39],[40], it implies that these five parameters can significantly influence the burstiness of the aggregated interference in multi-cell MIMO cellular networks. In Fig.~8, we analyze the impact of Nakagami shaping factor $m$ on the proposed interference model. Based on the PDF shape change in Fig. 8, it implies that the Nakagami shaping factor has little influence on the power of aggregated interference in multi-cell MIMO cellular networks.

\section{Capacity of Multi-cell Mimo Cooperative Cellular Networks}
\subsection{Cooperative System}
Based on the proposed interference model, we further investigate the capacity of a multi-cell MIMO
cooperative cellular network. The multi-cell MIMO cellular network considered in this paper is as
follows: in the two dimensional given region, there are infinite BSs with $N_t$ antennas and user
terminals with $N_r$ antennas, moreover the location of interfering BSs is assumed to follow the Poisson
spatial distribution. Every cell only has one BS. Without loss of generality, arbitrary three adjacent
BSs are selected for cooperative transmission and these three adjacent BSs are located in a triangle
structure. Furthermore, every BS is assumed to have three sectors and three adjacent sectors in
different BSs is configured a hexagon cooperative cluster structure as in Fig. 9. Therefore, these
three adjacent BSs are called the cooperative BSs and these three adjacent sectors are called a
cooperative cluster. The cooperative transmission is limited in the overlapping areas of the cells
and this area is called the cooperative transmission area. The rest of the area is called the
traditional transmission area. The detailed cooperative cluster structure is illustrated in Fig. 10.
To simplify the analysis, our capacity analysis of multi-cell MIMO cellular networks is limited to a
cooperative cluster.
Considering that the interference has great impact on the cell-edge users in the multi-cell MIMO
cellular network, this paper focuses on the downlink capacity of cell-edge users in the cooperative
transmission area. In the cell-edge overlapping area, $K$ users are assumed to simultaneously receive the
expected signal from the cooperative BSs $N_b\in[1,3]$ and the number of other users in this multi-cell MIMO cellular network is denoted as $K^{'}$. In the following numerical analysis, different numbers of cooperative BSs $N_b$
correspond to different cooperative transmission solutions.

\subsection{Downlink Capacity of Multi-cell MIMO Cooperative Cellular Network}
For analytical tractability, all wireless channels in the multi-cell MIMO cellular network are
assumed to be Nakagami-m fading channels with path loss. This assumption has been adopted in many
wireless communication studies (e.g., [41]--[45]) and a recent study [46] has shown that shadowing
effect does not cause major changes in the capacity PDF. Without loss of generality, the signal $y_k$
received by the cell-edge user $\mathrm{UE}_k$ is expressed as follows [26]
\begin{equation}
y_k=\sum_{b=1}^{N_b}\textbf{h}_{b,k,c}\textbf{x}_{b,k,c}
+\sum_{b=1}^{N_b}\textbf{h}_{b,k,c}\sum_{j=1,j\neq k}^{K}\textbf{x}_{b,j,c}
+\sum_{b^{'}=1}^{\infty}\textbf{h}_{b^{'},k,\overline{c}}\sum_{j=1}^{K^{'}}\textbf{x}_{b^{'},j,\overline{c}}
+\textbf{n}_0.   \tag{9}
\end{equation}
On the right side of (9), the first term is the expected signal for the $\mathrm{UE}_k$,
the second term is the aggregated interference signal from the cooperative cluster $c$,
the third term is the aggregated interference signal from the cluster $\overline{c}$,
where $\overline{c}$ is the complement cluster of the cluster $c$, i.e., one of all other clusters excluding the
cooperative cluster $\overline{c}$, and the fourth term $\textbf{n}_0$ is the additive white Gaussian noise (AWGN) in the
wireless channel. Where $\textbf{h}_{b,k,c}$ is the channel matrix between the BS $b$ and the user $\mathrm{UE}_k$ in the cooperative
cluster $c$, $\textbf{x}_{b,k,c}$ is the signal vector from the BS $b$ to the user $\mathrm{UE}_k$ in the cooperative cluster $c$, $\textbf{h}_{b^{'},k,\overline{c}}$ is
the channel matrix between the BS $b^{'}$ and the user $\mathrm{UE}_k$ in the cluster $\overline{c}$, $\textbf{x}_{b^{'},k,\overline{c}}$ is
the signal vector from the BS $b^{'}$ to the user $\mathrm{UE}_k$ in the cluster $\overline{c}$.

Considering the maximum rate transmission / maximum rate combining (MRT/MRC) approaches used in
the multi-cell MIMO cellular network [46]--[48], the signal-to-interference-and-noise ratio (SINR) received
by the user $\mathrm{UE}_k$ can be expressed by
\[\mathrm{SIN}{\mathrm{R}_k} = \frac{{{P_{\mathrm{ant}}}{\lambda _{\max }}\left( {{{\bf{H}}_{k,c}}{\bf{H}}_{k,c}^H} \right)}}{{{N_0} + \sum\limits_{b = 1}^\infty  {{I_b}} }}
\tag{10a}
\]
with \\
\begin{equation}
\textbf{H}_{k,c}=\left[\textbf{h}_{1,k,c}, \textbf{h}_{2,k,c}, \cdots,\textbf{h}_{N_{b},k,c}                               \right]
  \tag{10b}
\end{equation}
where $P_{\mathrm{ant}}$ and $I_{b}$ are the transmission power of each antenna and the sum of the interference signal
power transmitted by the $N_t$ interference sub-streams from the same BS $b$, respectively; $\textbf{H}_{k,c}$ is the cooperative channel matrix in cooperative cluster $c$,
which is composed by the sub-channel matrix from cooperative BSs to the user $\mathrm{UE}_k$, $\lambda_{max}\left(\textbf{AA}^{H}\right)$
is the maximum singular value of the matrix $\textbf{AA}^{H}$.

Based on the Shannon theory, the capacity of the interference channel linking user $\mathrm{UE}_k$ in the multi-cell MIMO cellular
network can be expressed by
\begin{align}
 C_k &= {\mathrm{B}_\mathrm{w}}{\log _2}\left[ {1 + \frac{{{P_{\mathrm{ant}}}{\lambda _{\max }}\left( {{{\bf{H}}_{k,c}}{\bf{H}}_{k,c}^H} \right)}}{{{N_0} + {P_{\mathrm{ant}}}\sum\limits_{b = 1}^\infty  {\frac{1}{{r_b^{{\sigma _r}}}}} {w_b}\left( {\sum\limits_{i = 1}^{{N_r}} {\sum\limits_{j = 1}^{{N_t}} {{{\left| {z_{b,i,j}^{}} \right|}^2}} } } \right)}}} \right]
\tag{11}
\end{align}
where $\mathrm{B}_\mathrm{w}$ is the bandwidth in the wireless link and $\mathcal{I}$ is the unit matrix. Considering that the
power of AWGN $N_0$ can be ignored when it compares to the power of the received interference signal [29],
the average capacity of multi-cell MIMO cellular network can be given by
\begin{align}
 C_{Aver} &\simeq \mathrm{B}_\mathrm{w} \int_{0}^{\infty}\log_2   \left(1+r_{b}^{-\sigma_{r}}\eta  \right)f(\eta)\mathrm{d}\eta  \tag{12a}
\end{align}
with \\
\begin{equation}
\eta = \frac{S_d}{S_I} =\frac{{{P_{\mathrm{ant}}}{\lambda _{\max }}\left( {{{\textbf{H}}_{k,c}}{\textbf{H}}_{k,c}^H} \right)}}
{{{P_{\mathrm{ant}}}\sum\limits_{b = 1}^\infty  r_b^{-\sigma_{r}} {w_b}\left( {\sum\limits_{i = 1}^{{N_r}} {\sum\limits_{j = 1}^{{N_t}} {{{\left| {z_{b,i,j}^{}} \right|}^2}} } } \right)}}
 \tag{12b}
\end{equation}
where $S_{d}$ and $S_{I}$ are the powers of the expected signal and aggregated interference signal, respectively.
When the power of expected signal and aggregated interference signal are assumed to be statistically
independent, the joint PDF of $S_{d}$ and $S_{I}$ is derived as
\begin{equation}
f(\eta)=\int_{0}^{\infty}f_{P^{Rx}}(z)f_{d}(\eta z)z \mathrm{d} z   \tag{13}
\end{equation}
where $f_{P^{Rx}}(z)$ is the PDF of aggregated interference signal and $f_{d}(\eta z)$ is the PDF of expected signal.
Considering an urban macro-cell with rich scattering environment, i.e., $\sigma_{r}=4$,
let us substitute (8a) and (13) into (12a). Ultimately, a new exact downlink average capacity of multi-cell
MIMO cellular network with co-channel interference is derived as
\begin{equation}
C_{Aver}=\sqrt{\frac{\gamma^{2}}{2\pi}}\int_{0}^{\infty}\log_{2}
          \left(
               1+r_{b}^{-\sigma_{r}} \eta
          \right)
\left(
      \int_{0}^{\infty}
         e^{-\gamma^{2}/2z}
         z^{-1/2}
         f_d(\eta z)
         \mathrm{d} z
\right)
\mathrm{d} \eta                           \tag{14}
\end{equation}
where $\gamma$ is given by (8b).

\subsection{Closed-form Downlink Capacity of Multi-cell MISO Cellular Network}
Because it is easier to integrate multiple antennas into the BSs than the user terminals
in practice, in this section we further derive a new closed-form downlink average capacity for
cell-edge users in a multi-cell MISO cooperative cellular network with co-channel interference.

In a cooperative cluster, one user can simultaneously receive the desired signals and interfering
signals transmitted from cooperative BSs. To simplify our derivation, different users in the same coopeartive cluster are assumed to have orthogonal channels, i.e., the multi-user interference within a single cooperative cluster is assumed to be negligible.
Moreover, the power transmitted by every antenna in one BS is normalized to 1, i.e., $P_{\mathrm{ant}}=1$. To calculate the downlink capacity of multi-cell MISO cellular network,
we should first get the covariance of the desired signal, which has been derived in the Appendix and is expressed
by
\[{\textbf{R}_{xx}} = {P_{\mathrm{ant}}}\sum\limits_{b = 1}^{{N_b}} {\sum\limits_{j = 1}^{N_t^c} {\frac{1}{{r_b^{{\sigma _r}}}}{{\left| {z_{b,j}^{}} \right|}^2}} }  = \sum\limits_{b = 1}^{{N_b}} {\sum\limits_{j = 1}^{N_t^c} {\frac{1}{{r_b^{{\sigma _r}}}}{{\left| {z_{b,j}^{}} \right|}^2}} }  \tag{15}
\]
where $N_t^c$ is the number of cooperative antennas transmitting the desired signal and $r_b$ is the distance between the cooperative BS $b$ and a cell-edge user.

For a single cell MISO cellular network, the downlink capacity of user $\mathrm{UE}_k$ is typically presented
as follows [49]
\begin{equation}
C_{k}= \mathrm{B}_\mathrm{w}\log_2
     \left|
     \mathcal{I}
     +\frac
           {\textbf{R}_{xx}}
           {N_{0}+{P^{{R_X}}}}
     \right|
\tag{16}
\end{equation}
where $P^{{R_X}}$ is the aggregated interference.

Based on the assumption in the Appendix, the interfering signals and the desired signals from
different cooperative BSs in the cooperative cluster $c$ to the user $\mathrm{UE}_k$ are assumed to be uncorrelated.
Ideally, the interference within the cooperative cluster can be cancelled completely. Therefore,
from the proposed interference model in (2), the aggregated interference in
a multi-cell MISO cellular network is further expressed as follows
\begin{equation}
{P^{{R_X}}}= P_{\mathrm{ant}}
 \sum_{i=1}^{\infty}
 \sum_{j=1}^{N_t}
    \frac
       {1}{r_i^{\sigma_{r}}}
       \left|
       z_{i,j}
       \right|^2
       =
       \sum_{i=1}^{\infty}
 \sum_{j=1}^{N_t}
    \frac
       {1}{r_i^{\sigma_{r}}}
       \left|
       z_{i,j}
       \right|^2
\tag{17}
\end{equation}
where $r_i$ is the distance between the interfering BS  $i$ and the cooperative cluster and the locations of interfering BSs are governed by the Poisson spatial distribution, $z_{i,j}$ is
a random variable following the Nakagami-m distribution and represents a signal passing through
the Nakagami-m fading channel from antenna $j$ at BS $i$.

Furthermore, we substitute (15) and (17) into (16). Considering the path loss coefficient $\sigma_r=4$ in
the urban macro-cell with rich scattering environment, the new downlink capacity of multi-cell
MISO cooperative cellular network with co-channel interference is given by
\[{C_k} = \mathrm{B}_\mathrm{w}{\log _2}\left( {1 + \frac{{\sum\limits_{b = 1}^{{N_b}} {\sum\limits_{j = 1}^{N_t^c} {\frac{1}{{r_b^4}}{{\left| {z_{b,j}^{}} \right|}^2}} } }}{{{N_0} + \sum\limits_{i = 1}^\infty  {\sum\limits_{j = 1}^{{N_t}} {\frac{1}{{r_i^4}}{{\left| {z_{i,j}^{}} \right|}^2}} } }}} \right).
\tag{18}
\]

In this paper, we focus on cell-edge users located in cooperative areas of multi-cell MISO
cellular network. Therefore the distances $r_b$ between cooperative BSs and cell-edge users can be
approximated as equivalent. Furthermore, the path loss between cooperative BSs and users is
approximated as a constant. To simplify the notation, we define
\begin{equation}
\eta'=\frac{S_d^{'}} {S_I^{'}}= \frac{\sum_{b=1}^{N_b}\sum_{j=1}^{N_t^c}\left|z_{b,j}\right|^2} {\sum_{i=1}^{\infty}\sum_{j=1}^{N_t}\frac{1}{r_i^{4}}\left|z_{i,j}\right|^2}.
\tag{19}
\end{equation}
Considering that the power of AWGN $N_0$ can be ignored when it compares to the power of the received
interference $P^{{R_X}}$ [29], (18) can be concisely expressed as
\begin{equation}
{C_k} \approx \mathrm{B}_\mathrm{w}\log_2
  \left(
      1+r_{b}^{-4}\eta'
  \right).
\tag{20}
\end{equation}

The Nakagami shaping factor $m$ is configured as $m=1$ for both desired signals
and interfering signals. Each signal passing through the Nakagami fading channel
is assumed to follow an i.i.d complex Gaussian distribution with zero mean and variance of 1, i.e.,
\mbox{$z_{b,j} \sim CN(0,1)$}. Thus, the PDF of aggregated expected signal
$S_d^{'}$ can be simplified as
\begin{equation}
f_d(x)=\frac
       {x^{N_b N_t^c -1} e^{-x/2}}
       {(N_b N_t^c -1)!2^{N_b N_t^c}}
       , x>0
\tag{21}.
\end{equation}

On the other hand, the PDF of aggregated interference $S_I^{'}$ with complex channel
coefficients has already been derived in (7a). We assume that the interference signal passes
through the same channel condition of the desired signal, i.e. the Nakagami-m fading channels
with path loss, thus the PDF of the aggregated interference $S_I^{'}$ can be further
simplified as
\begin{equation}
f_{P^{Rx}}(y) = \sqrt{\frac {\gamma^{2}} {2\pi}}\frac{e^{-\gamma^{2}/(2y)}}{y^{3/2}}, y>0   \tag{22a}
\end{equation}
with \\
\begin{equation}
\gamma=
      \frac
      {2
      \Gamma
      \left(
      \frac{3}{2}
      \right)
      \pi \lambda_{BS}
      \Gamma(N_t+\frac{1}{2})
      }
      {(N_t-1)!}.
\tag{22b}
\end{equation}

In this paper, the expected signal $S_d^{'}$ and the aggregated interference $S_I^{'}$ are
assumed statistically independent. Therefore the PDF of $\eta^{'}$ is given by
\begin{align}
f(\eta')&=
\frac
  {\sqrt{\frac {\gamma^{2}} {2\pi}}}
  {(N_b N_t^c -1 )! 2^{N_b N_t^c}}
  \eta^{'N_b N_t^c -1}
  \int_{0}^{\infty}
\mathrm{exp}
  \left(
  -\frac{\gamma^{2}}{2z}
  -\frac{\eta 'z}{2}
  \right)
  z^{v-1} \mathrm{d} z.
\tag{23}
\end{align}
Based on the table of integral in [37], (23) can be simplified as
\begin{equation}
f(\eta ')=
{\frac{\gamma^{v+1}\sqrt{2/\pi}}
{(N_b N_t^c -1)! 2^{N_b N_t^c}}}\eta'^{\frac{v-1}{2}}
K_{v}
\left(
\gamma \sqrt{\eta'}
\right)
\tag{24a}
\end{equation}
with \\
\begin{equation}
v=N_b N_t^c-\frac{1}{2}.
\tag{24b}
\end{equation}

Based on (24a)--(24b), the average channel capacity can be derived by calculating the expectation of (20). After normalizing the bandwidth $\mathrm{B}_\mathrm{w}=1$,
the normalized downlink average capacity of multi-cell MISO cooperative cellular network
with co-channel interference is derived as
\begin{equation}
\overline{C}_{Aver} =\frac{C_{Aver}}{\mathrm{B}_\mathrm{w}}
=p\int_{0}^{\infty}
\log_2
   \left(
   1+r_b^{-4}\eta'
   \right)
   \eta'^{\frac{v-1}{2}}
   K_v
   \left(
   \gamma \sqrt{\eta'}
   \right)
   \mathrm{d} \eta'
\tag{25a}
\end{equation}
with \\
\begin{equation}
p=\frac{\gamma^{v+1}\sqrt{2/\pi}}
{(N_b N_t^c -1)! 2^{N_b N_t^c}}
\tag{25b}
\end{equation}
where $v$ and $\gamma$ are given by (24a) and (22b), respectively.

The integral in (25a) will cause some difficulties in the practical engineering calculation. To simplify the expression, we will utilize a special function, the so-called Meijer's G-function, which is defined as [50]
\[
G_{p,q}^{m,n} \left( {x\left| \begin{array}{l}
 a_1 , \cdots a_p  \\
 b_1 , \cdots b_q  \\
 \end{array} \right.} \right) = \frac{1}{{2\pi \Im }}\int_0^\infty
 {\frac{{\prod\limits_{j = 1}^m {\Gamma (b_j  - s)}
 \prod\limits_{j = 1}^n {\Gamma (1 - a_j  + s)} }}
 {{\prod\limits_{j = m + 1}^q {\Gamma (1 - b_j  + s)}
 \prod\limits_{j = n + 1}^p {\Gamma (a_j  - s)} }}x^s \mathrm{d}s}
\tag{26}
\]
where $0 \le m \le q$, $0 \le n \le p$, and $\Im  = \sqrt { - 1}$. The $\log _2 \left(  \cdot \right)$ and
$K_v \left(  \cdot  \right)$ functions can be expressed as a special form of Meijer's G-function. So, (25a) can be expressed by a new form with Meijer's G-functions
\[
\bar C_{Aver}  = \frac{p}{{2\mathrm{ln}2}}\int_0^\infty  {\eta'^{\frac{{v - 1}}{2}} G_{2,2}^{1,2}
\left( {r_b^{ - 4} \eta ' \left| {\begin{array}{*{20}c}
   {1,1}  \\
   {1,0}  \\
\end{array}} \right.} \right)} G_{0,2}^{2,0} \left( {\begin{array}{*{20}c}
   {}  \\
   {\frac{v}{2}, - \frac{v}{2}}  \\
\end{array}\left| {\frac{{\gamma ^2 \eta ' }}{4}} \right.} \right)\mathrm{d}\eta'.
\tag{27}
\]

Based on the table of integral in [37], (25a) can be further simplified as follows by eliminating the integral calculation and parameter
simplification
\[
\bar C_{Aver}  = \frac{p}{{2\mathrm{ln}2}}r_b^{2(v + 1)} G_{2,4}^{4,1} \left( {\begin{array}{*{20}c}
   { - \frac{{v + 1}}{2},\frac{{1 - v}}{2}}  \\
   { - \frac{v}{2},\frac{v}{2}, - \frac{{v + 1}}{2}, - \frac{{v + 1}}{2}}  \\
\end{array}\left| {\frac{{\gamma ^2 r_b^4 }}{4}} \right.} \right)
\tag{28}
\]
where $p$, $v$ and $\gamma$ are given by (25b), (24a) and (22b), respectively.

\section{Numerical Results And Discussion For Miso Capacity Model}

Based on the normalized downlink average capacity of multi-cell MISO cellular
network with co-channel interference derived in Section III, the effect of various system
parameters on the capacity will be analyzed numerically in this section.
In our numerical analysis, some parameters of the capacity model are configured as follows: $\sigma _{dB}  = 7$, $m=1$, $\sigma _r  = 4$, $\sigma ^2  = 1$,
$\lambda _{BS}  = 1/(\pi  * 500^2 )$, and $r_b = 500$ m. Every BS has no more than four antennas used for cooperative transmission,
but in the default cooperative transmission scheme, every cooperative BS just uses two antennas. To simplify numerical analysis, the cooperative
transmission among BSs is assumed as the aggregation of desired signals transmitted
by the multi-antennas from different BSs, i.e., desired signals are accumulated directly by numerical calculation. The detailed cooperative transmission scheme, e.g., the joint
pre-coding scheme [25], is not included in this paper for conciseness.

In Fig. 11, we first analyze the impact of the number of cooperative transmission (Co-Tx)
antennas per cooperative BS on the normalized downlink average capacity. The number of
cooperative BSs is noted as $CBS$. Without cooperative BSs, i.e., $CBS=1$, the capacity is improved by 209\% when the number of Co-Tx antennas per cooperative BS is increased
from 1 to 4. With two cooperative BSs, i.e., $CBS=2$, the capacity is improved by
173\% when the number of Co-Tx antennas per cooperative BS is increased from 1 to 4.
With three cooperative BSs, i.e., $CBS=3$, the capacity is improved by 153\%
when the number of Co-Tx antenna per cooperative BS is increased from 1 to 4. When only
one antenna per cooperative BS is used to transmit expected signals, the capacity
is improved by 80.99\% when CBS is increased
from 1 to 2; the capacity is improved by 37.88\% when CBS is increased from 2 to 3. When four antennas per cooperative BS is used to
transmit expected signals, the capacity is improved by 50.9\% when CBS is increased from 1 to 2; the capacity is
improved by 27.62\% when CBS is increased from 2 to 3.
Therefore, with the increasing number of antennas or
cooperative BSs, the capacity performance is improved, but the increment of capacity performance is decreased.

In Fig. 12, the impact of the density parameter of interfering BSs $\lambda _{BS}$ on the normalized
downlink average capacity is analyzed by numerical simulations. Without cooperation BSs,
i.e., $CBS=1$, the capacity decreases by 94.86\% when the density parameter of interfering BSs is
increased from $0.5 \times 10^{ - 6} {\rm{m}}^{{\rm{ - 2}}}$ to $3.5 \times 10^{ - 6} {\rm{m}}^{{\rm{ - 2}}}$.
In the case of two cooperative BSs, i.e., $CBS=2$, the capacity decreases by 93.40\% when
the density parameter of interfering BSs is increased
from $0.5 \times 10^{ - 6} {\rm{m}}^{{\rm{ - 2}}}$ to $3.5 \times 10^{ - 6} {\rm{m}}^{{\rm{ - 2}}}$.
In the case of three cooperative BSs, i.e., $CBS=3$, the capacity decreases by 92.22\%
when the density parameter of interfering BSs is increased from $0.5 \times 10^{ - 6} {\rm{m}}^{{\rm{ - 2}}}$
to $3.5 \times 10^{ - 6} {\rm{m}}^{{\rm{ - 2}}}$. These results
indicate that the density parameter of interfering BSs can obviously decrease the capacity
performance in a multi-cell MISO cellular network. When the density parameter of
interfering BSs is configured as $0.5 \times 10^{ - 6} {\rm{m}}^{{\rm{ - 2}}}$,
compared with the capacity of one BS,
the average capacity of two cooperative BSs improves by 48.76\%; compared with the average capacity
of two cooperative BSs, the average capacity of three cooperative BSs improves by 21.99\%. Hence,
with the specified density of interfering BSs, the cooperative transmission can improve
the capacity performance, but this gain of capacity performance is decreased with the
increasing number of cooperative BSs. When the density of interfering BSs is configured
as $3.5 \times 10^{ - 6} {\rm{m}}^{{\rm{ - 2}}}$,
compared with the capacity of one BS, the average capacity of two cooperative BSs improves by 90.98\%; compared with the average capacity of two cooperative BSs, the average capacity
of three cooperative BSs improves by 43.81\%. Therefore, compared with
the capacity performance in the small density of interfering BSs, the cooperative
transmission can obviously improve the capacity performance with a high density of
interfering BSs.

\section{Conclusions}

In this paper, we have derived the exact downlink average capacity of multi-cell MIMO
cellular network with co-channel interference. Furthermore, the analytical closed-form
normalized downlink average capacity for cell-edge users in a multi-cell MISO cooperative
cellular network with co-channel interference has been derived and analyzed numerically.
To derive this downlink capacity model, an analytical co-channel interference model has been proposed for multi-cell MIMO cellular networks. Based on the proposed closed-form normalized downlink average capacity
of multi-cell MISO cooperative cellular network with co-channel interference, numerical
results have shown that the cooperative transmission can improve the capacity performance in most
cases, but the capacity gains suffer from the increased number of cooperative BSs or antennas.
Our analysis indicates that the cooperative transmission can efficiently enhance the
capacity performance, especially in scenarios with high densities of interfering BSs.
For our future work, we will explore the impact of different cooperative transmission
schemes on the system capacity.

\section*{Appendix: Derivation of (15)}
In this appendix, we derive the covariance of expected signals at the user $\mathrm{UE}_k$.
In one cooperative cluster, one cell-edge user can receive expected signals and
unexpected signals transmitted from different cooperative BSs. To simplify the complex of
derivation, expected signals and unexpected signals from
different cooperative BSs in the same cooperative cluster $c$ received by the user $\mathrm{UE}_k$ are
assumed to be uncorrelated, which can be expressed as follows
\[
\mathbb{E}\left[ {\textbf{x}_{b,k,c} \left( {\textbf{x}_{b',k,c} } \right)^H } \right] = 0,\forall b \ne b'
\tag{29}
\]
where $\mathbb{E}\left(  \cdot  \right)$ is the expectation calculation operator,
$\textbf{x}_{b,k,c}$ is the expected signal transmitted to
the user $\mathrm{UE}_k$ by BS $b$ in the cooperative cluster $c$, $\textbf{x}_{b^{'},k,c}$ is the unexpected signal transmitted
to the user $\mathrm{UE}_k$ by the BS $b^{'}$ in the cooperative cluster c.
Therefore, the covariance of the expected signal $\textbf{Q}_{b,k,c}$ can be given by
\[
\textbf{Q}_{b,k,c}  = \mathbb{E}\left[ {\textbf{x}_{b,k,c} \textbf{x}_{b,k,c}^H } \right] = P_{\mathrm{ant}} {\cal I}
\tag{30}
\]
where $\cal I$ is the unite matrix.
Furthermore, in the scenario of multi-cell MISO cooperative cellular network,
the signal covariance transmitted to the user $\mathrm{UE}_k$ by different BSs in the cooperative
cluster $c$ can be expressed by
\begin{align}
R_{xx} &= \mathbb{E}\left( {\sum\limits_{b = 1}^{N_b } {{\textbf{h}}_{b,k,c} \textbf{x}_{b,k,c}
\left( {\textbf{x}_{b,k,c} } \right)^H \left( {{\textbf{h}}_{b,k,c} }
\right)^H } } \right) +
\mathbb{E}\left( {\sum\limits_{b = 1}^{N_b }
{\sum\limits_{\scriptstyle b' = 1 \hfill \atop
  \scriptstyle b' \ne b \hfill}^{N_b }
  {{\textbf{h}}_{b,k,c} \textbf{x}_{b,k,c} \left( {\textbf{x}_{b',k,c} }
  \right)^H \left( {{\textbf{h}}_{b',k,c} } \right)^H } } } \right)
  \tag{31}
\end{align}
where ${\textbf{h}}_{b,k,c}$ and ${\textbf{h}}_{b^{'},k,c}$ are the channel matrices from the BS $b$ and BS $b'$ to the user $\mathrm{UE}_k$
in the cluster $c$, respectively.
Base on (29) and (30), we only consider the Nakagami-m fading effect and (31) can be further simplified as
\begin{align}
R_{xx}  &=
\mathbb{E}
\left( {\sum\limits_{b = 1}^{N_b } {{\bf{h}}_{b,k,c} \textbf{x}_{b,k,c}
\left( {\textbf{x}_{b,k,c} } \right)^H \left( {{\bf{h}}_{b,k,c} } \right)^H } } \right) \notag \\
&=
 \sum\limits_{b = 1}^{N_b } {\left[ {{\textbf{h}}_{b,k,c}
 \textbf{Q}_{b,k,c} \textbf{h}_{b,k,c}^H } \right]}   \notag \\
 &=P_{ant} \sum\limits_{b = 1}^{N_b } {\left\| {{\textbf{h}}_{b,k,c} } \right\|_F^2 } \notag \\
 &= P_{ant} \sum\limits_{b = 1}^{N_b }
 {\sum\limits_{j = 1}^{N_t^c } {\left| {{\textbf{h}}_{b,j}} \right|^2 } }    \notag \\
 &= P_{ant} \sum\limits_{b = 1}^{N_b }
  {\sum\limits_{j = 1}^{N_t^c } {\frac{1}{{r_b^{\sigma ^r } }}
  \left| {z_{b,j} } \right|^2 } }.  \tag{32}
\end{align}
This completes the derivation.

\newpage
\begin{figure}
\vspace{0.1in}
\centerline{\includegraphics[width=8cm,draft=false]{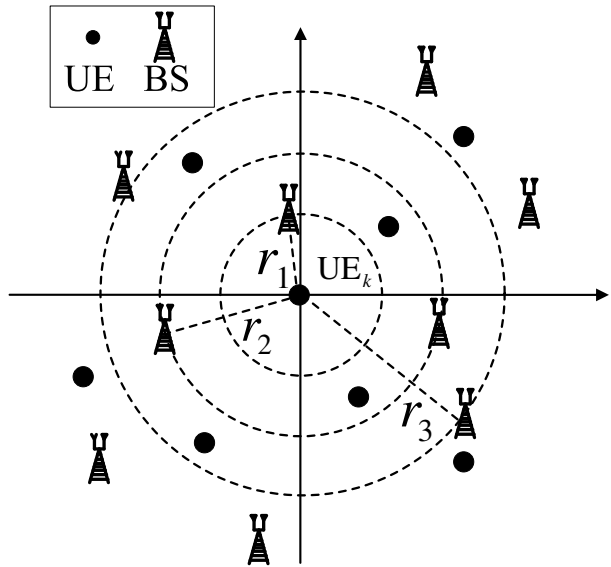}}
\caption{\small General interference model.}
\end{figure}

\begin{figure}
\vspace{0.1in}
\centerline{\includegraphics[width=13cm,draft=false]{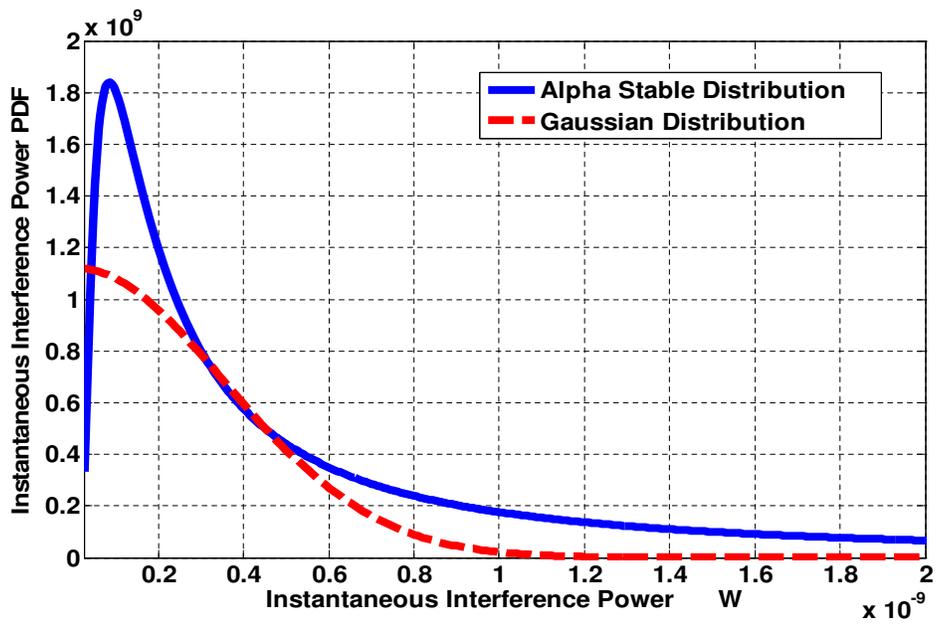}}
\caption{\small Comparison of interference power PDFs between the alpha-stable distribution and the Gaussian distribution.}
\end{figure}

\begin{figure}
\vspace{0.1in}
\centerline{\includegraphics[width=12cm,draft=false]{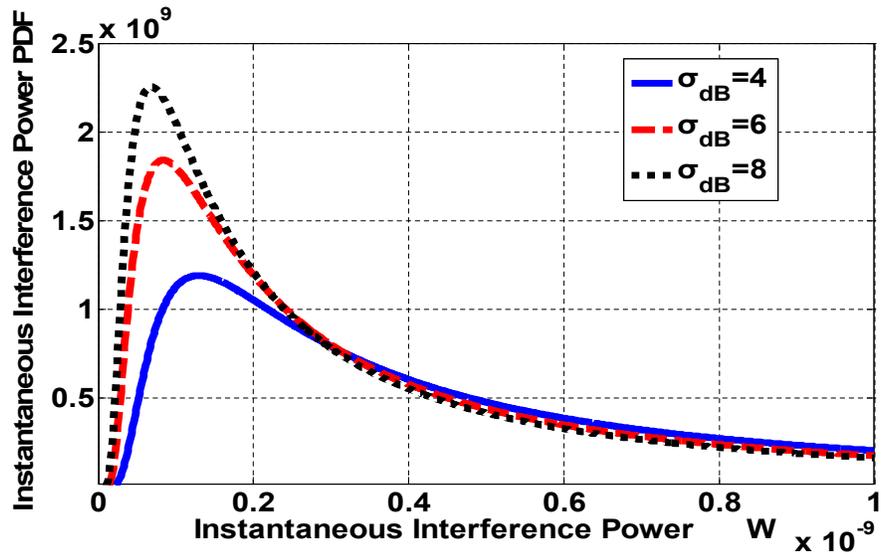}}
\caption{\small Impact of shadowing spread parameter $\sigma_{dB}$ on the PDF of the aggregated interference.}
\end{figure}

\begin{figure}
\vspace{0.1in}
\centerline{\includegraphics[width=12cm,draft=false]{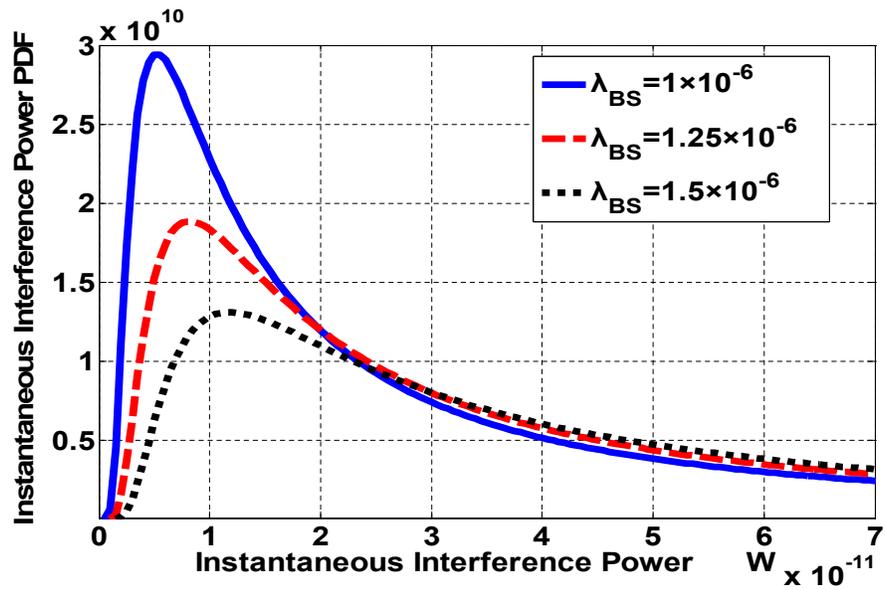}}
\caption{\small  Impact of base stations density parameter $\lambda_{BS}$ on the PDF of the aggregated interference.}
\end{figure}

\begin{figure}
\vspace{0.1in}
\centerline{\includegraphics[width=12cm,draft=false]{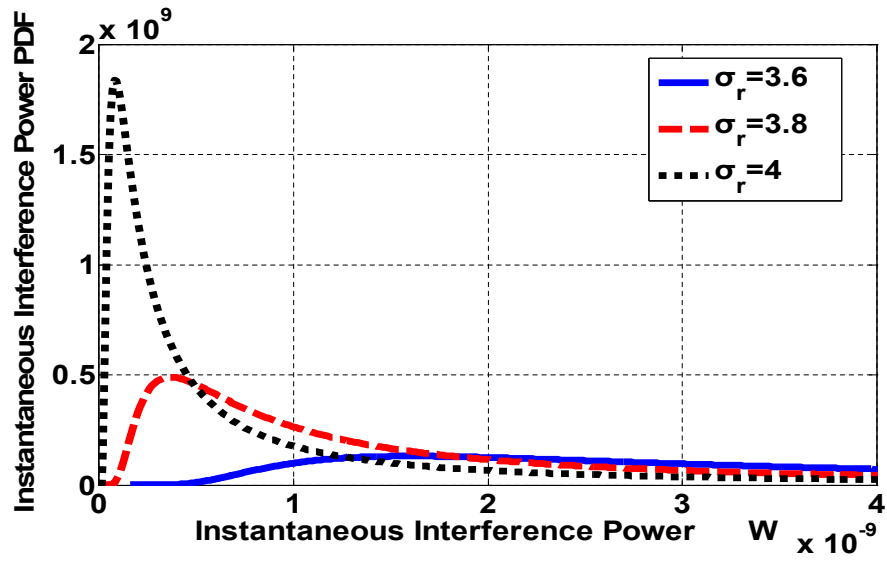}}
\caption{\small Impact of path loss coefficient parameter $\sigma_{r}$ on the PDF of the aggregated interference.}
\end{figure}

\begin{figure}
\vspace{0.1in}
\centerline{\includegraphics[width=13cm,draft=false]{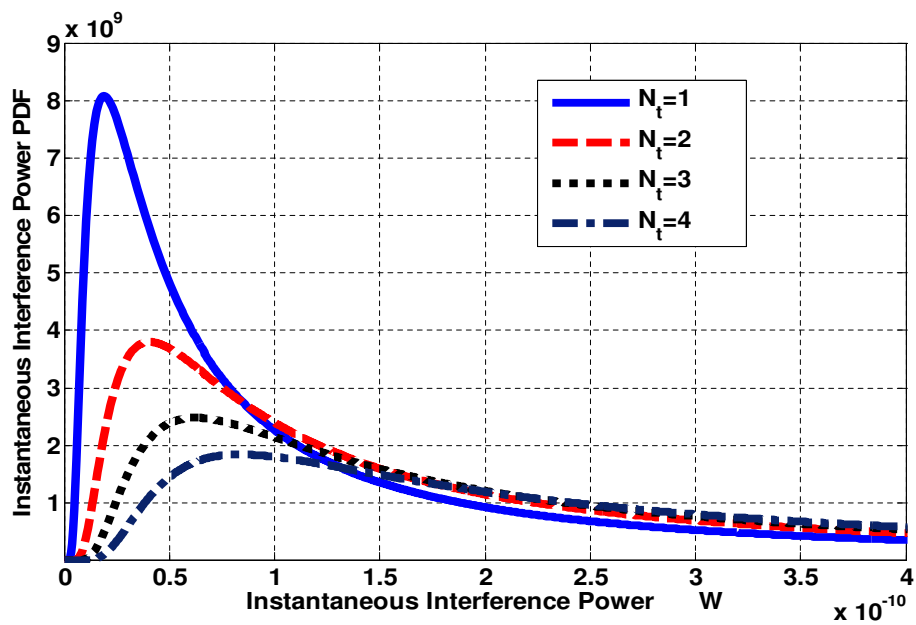}}
\caption{\small Impact of number of transmission antennas per interfering transmitter $N_t$ on the PDF of the aggregated interference.}
\end{figure}

\begin{figure}
\vspace{0.1in}
\centerline{\includegraphics[width=12cm,draft=false]{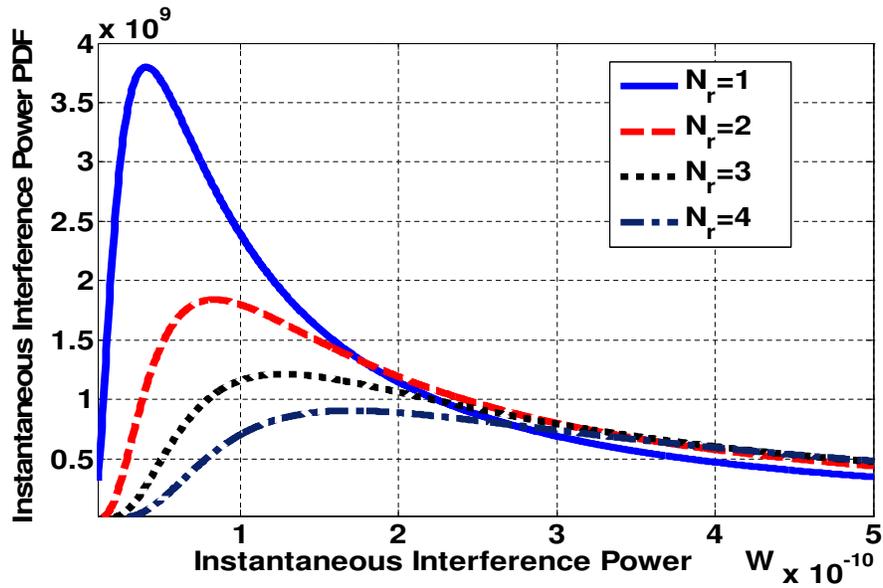}}
\caption{\small Impact of number of receiving antennas per user $N_r$ on the PDF of the aggregated interference.}
\end{figure}

\begin{figure}
\vspace{0.1in}
\centerline{\includegraphics[width=12cm,draft=false]{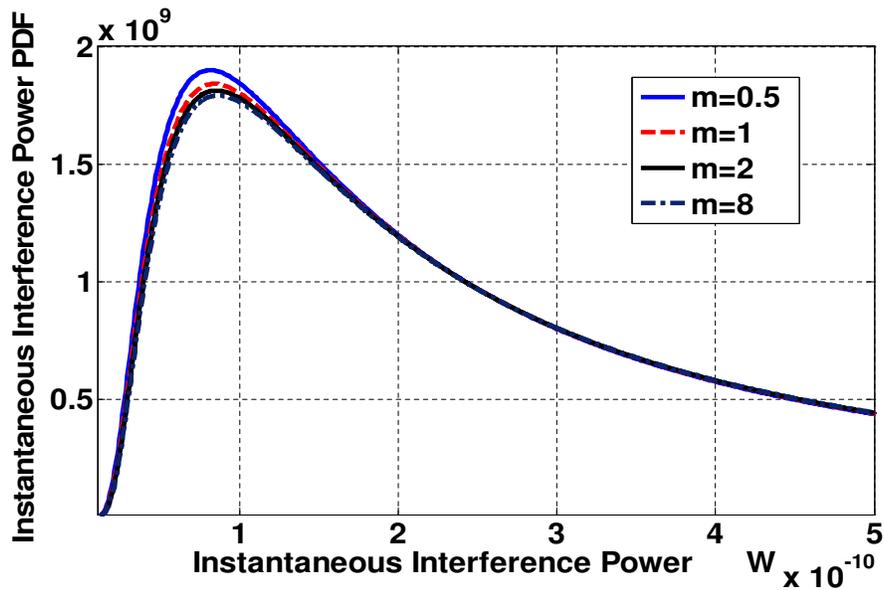}}
\caption{\small Impact of Nakagami shaping factor $m$ on the PDF of the aggregated interference.}
\end{figure}

\begin{figure}
\vspace{0.1in}
\centerline{\includegraphics[width=10cm,draft=false]{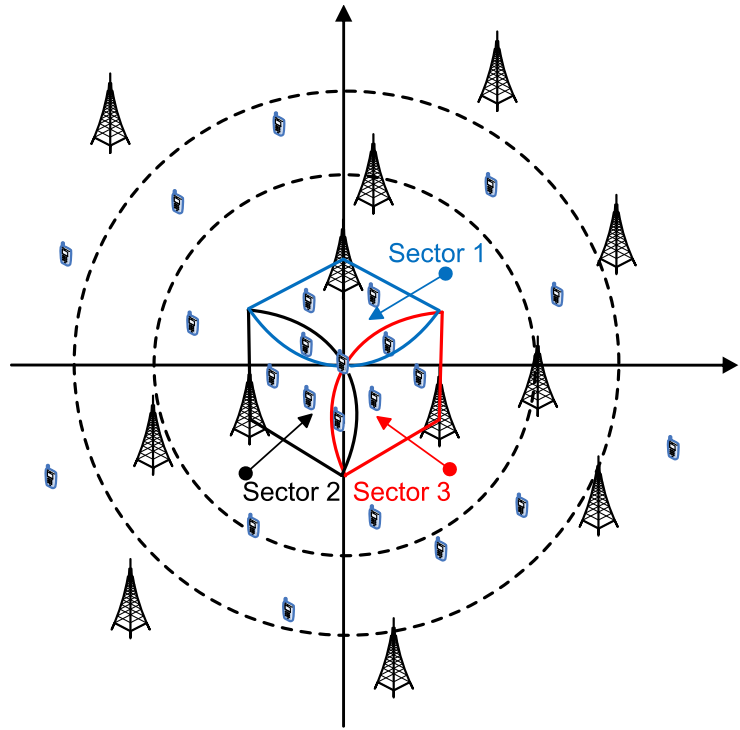}}
\caption{\small Cooperative system with co-channel interference.}
\end{figure}

\begin{figure}
\vspace{0.1in}
\centerline{\includegraphics[width=12cm,draft=false]{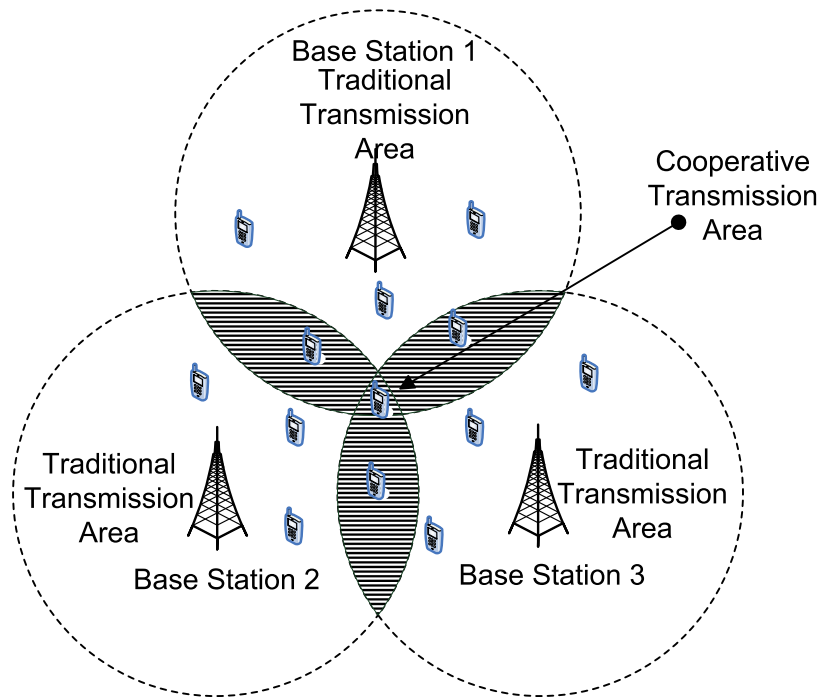}}
\caption{\small A cooperative cluster structure of the multi-cell MIMO cellular network.}
\end{figure}

\begin{figure}
\vspace{0.1in}
\centerline{\includegraphics[width=13cm,draft=false]{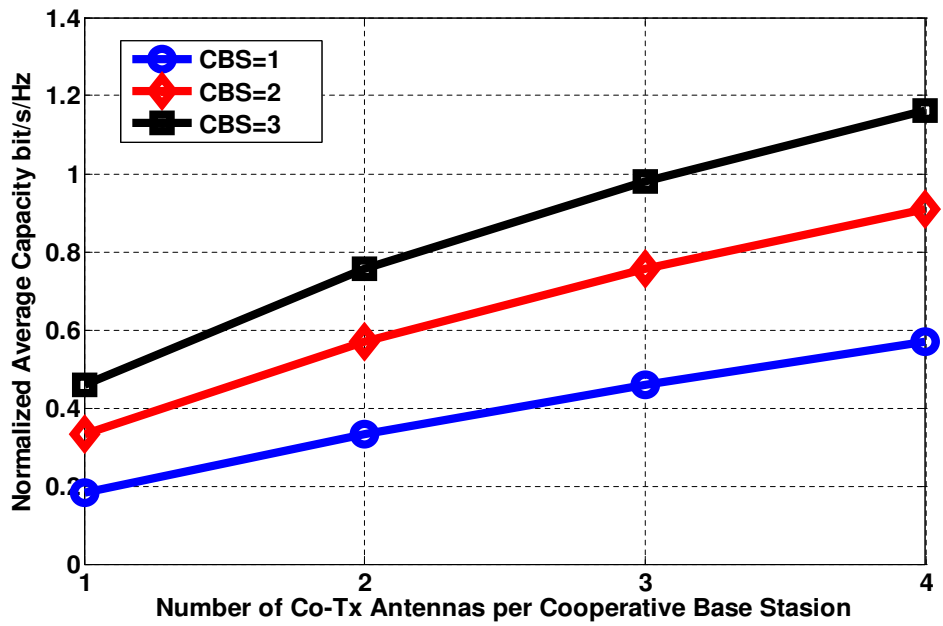}}
\caption{\small Impact of the number of Co-Tx antennas per cooperative BS on the normalized downlink average capacity of multi-cell MISO cellular networks.}
\end{figure}

\begin{figure}
\vspace{0.1in}
\centerline{\includegraphics[width=13cm,draft=false]{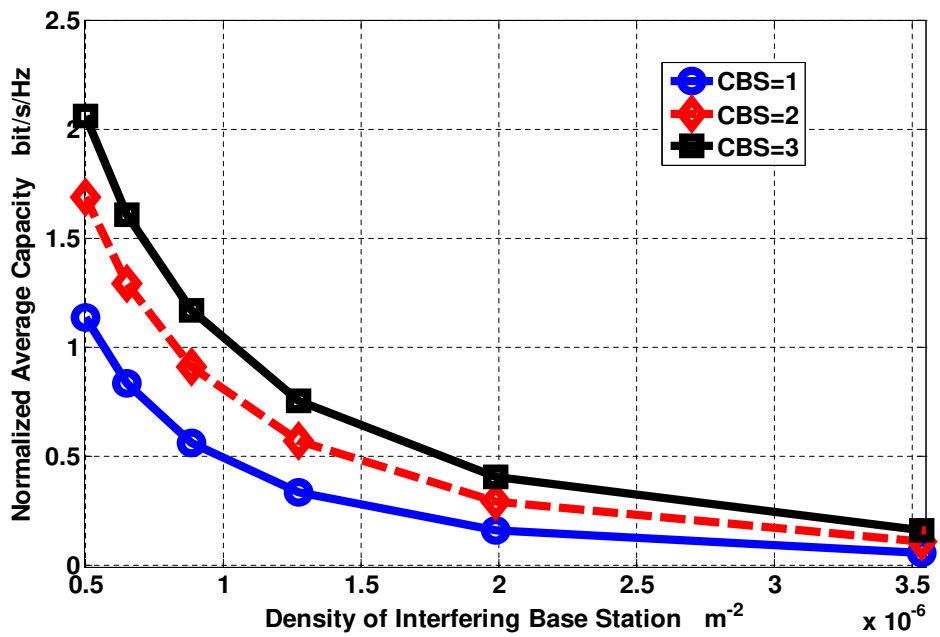}}
\caption{\small Impact of the density of interfering BSs on the normalized downlink average capacity of multi-cell MISO cellular networks.}
\end{figure}

\end{document}